\documentclass[aps,prl,floatfix,twocolumn,showpacs]{revtex4} 
\usepackage{graphicx} \usepackage{amsmath} \usepackage{amssymb} 
\newcommand{\BEQ}{\begin{equation}}
\newcommand{\EEQ}{\end{equation}}
\newcommand{\BEA}{\begin{eqnarray}}
\newcommand{\EEA}{\end{eqnarray}}

\renewcommand{\d}{{\rm d}}

\newcommand{\half}{\frac{1}{2}}

\begin{document} 
\draft
\title
{Minimal Work Principle and its Limits for Classical Systems.}
\date{\today}
\author{A.E. Allahverdyan$^{1)}$,
Th.M. Nieuwenhuizen$^{2)}$}
\address{
$^{1)}$Yerevan Physics Institute,
Alikhanian Brothers St. 2, Yerevan 375036, Armenia,\\
$^{2)}$ Institute for Theoretical Physics,
Valckenierstraat 65, 1018 XE Amsterdam, The Netherlands,
}

\begin{abstract}

The minimal work principle asserts that work done on a thermally
isolated equilibrium system, is minimal for the slowest
(adiabatic) realization of a given process.  This principle, one of the
formulations of the second law, is operationally well-defined for any
finite (few particle) Hamiltonian system. Within classical
Hamiltonian mechanics, we show that the principle is valid for a system
of which the observable of work is an ergodic function. For non-ergodic systems 
the principle may or may not hold, depending on additional conditions.
Examples displaying the limits of the principle are presented and their 
direct experimental realizations are discussed.

\end{abstract}



\maketitle

Thermodynamics originated in the nineteenth century as a science of
macroscopic machines, constructed for transferring, applying or
transmitting energy \cite{thermobooks}. The main results of this science
are summarized in several formulations of the second law
\cite{thermobooks}. Last fifty years witnessed progressive
miniaturization of the components employed in the construction of
devices and machines \cite{venturi}.  This will open the way to new
technologies in the fields of medicine, computation and renewable energy
sources.  

For microscopic machines and devices, we need to
understand how the second law applies to small systems.  There are two
aspects of this program: {\it i)} Emergence of the second law, where one
studies fluctuations of work or entropy, knowing that on the average they
satisfy the second law. Important contributions to this topic were
made by Smoluchowski and others, nearly hundred years ago \cite{epstein}.
Recently, the activity in this field revived in the context of
fluctuation theorems \cite{kur}. {\it ii)} Limits of the second law,
where the very formulation 
is studied from first principles.  
Here we continue on that and study (limits of) the minimal work principle based
on classical mechanics.  This formulation of the second law is
operationally well-defined for finite systems, and it relates the energy
cost of an operation to its speed.  The principle was deduced
from experience and postulated in thermodynamics, where it is equivalent
to other formulations of the second law.  Its derivation in statistical
physics was formulated several times, on various levels of generality
\cite{deri,deri1,minima}. Almost all these studies concentrate on
macroscopic systems and confirm the validity of the principle. 
In Ref.~\cite{minima} the principle was derived for a
finite quantum systems, and limits, related to energy level crossing,
were indicated. However, the reduction level of quantum mechanics
is not always needed, e.g., certain aspects of nanoscience
are adequately understood already with classical ideas \cite{venturi}.
In addition, it is not easy to design experiments in the quantum
domain that would check the validity of formulations of the
second law.  Thus it is necessary to understand the principle in 
classical mechanics and this is our present purpose. Our results will
apply to small systems, having one degree of freedom. 

Consider a classical system with $N$ degrees of freedom and 
with a Hamiltonian $H(q,p,R_t)$, where
$q=(q_1,...,q_N)$ and $p=(p_1,...,p_N)$ are, respectively, canonical
coordinates and momenta.
The interaction with external sources of work is realized via a
time-dependent parameter $R_t=r(t/\tau_R)$ with time-scale $\tau_R$.  
The motion starts at the initial time $t_{\rm
i}=\bar{t}_{\rm i}\tau_R$, and we follow it till the final time $t_{\rm
f}=\bar{t}_{\rm f}\tau_R$.  We shall denote $z\equiv (q,p)$, $\d
z\equiv\d q\, \d p$ and $H_{\rm k}=H(z,R_{\rm k})$ with ${\rm k}={\rm
i}$, $\rm f$, for the initial and final values, respectively.  

The system starts its evolution from an equilibrium Gibbs state at
temperature $T={1}/{\beta}>0$: ${\cal P}_{\rm i}(z)=e^{-\beta
H_{\rm i}(z)}/{\cal Z}_{\rm i}$, 
where ${\cal Z}_{\rm i}=\int\d z\,e^{-\beta H_{\rm i}(z)}$. 
The work done on the system is the average energy difference \cite{thermobooks}
\BEA
\label{wo}
W=\int \d z[{\cal P}_{\rm f}(z) H_{\rm f}(z)-{\cal P}_{\rm i}(z) H_{\rm i}(z)]\\
=\int_{t_{\rm i}}^{\rm t_f}\d s\, \dot{R}_s\, {\cal P}(z,s)\,  \partial_R H(z,R_s),
\label{wo1}
\EEA
where ${\cal P}_{\rm f}(z)$ is the final distribution, and where
the equivalence between (\ref{wo}) and (\ref{wo1}) is established with help of 
the Liouville equation for ${\cal P}$ and the standard boundary condition ${\cal P}(z)=0$ 
for $z\to\pm\infty$ \cite{thermobooks}. 
Eq.~(\ref{wo1}) justifies  to call $w(z,R)\equiv\partial_RH(z,R)$
``the observable of work''. 
Consider processes with different $\tau_R$, but the same
$r(\bar{t})$ and $R_{\rm k}=r(\bar{t_{\rm k}})$ (${\rm k}={\rm i, f}$).
{\it The minimal work principle} claims
\BEA
\label{dwo}
\Delta W\equiv W- \widetilde{W}\geq 0,
\EEA
where $\widetilde{W}$ is the work for the adiabatically slow
realization, where $\tau_R$ is much larger than the characteristic time
$\tau_S$ of the system.  
Eq.~(\ref{dwo}) is an optimality statement: 
the smallest amount of work to be put into the system ($W>0$) is the adiabatic one,
and the largest amount of work to be extracted from the system ($W<0$) is again
the adiabatic one. 
The condition of being slow is purely
operational.  If there exist several widely different characteristic
times, there are several senses of being slow. 

Our derivation of the principle consists of three steps and
assumes that $w(z,R)$ is an ergodic observable of the dynamics
with $R=$const. Note that the known inequality $W\geq F_{\rm f}-F_{\rm
i}$, where $F_{\rm f,i}=-T\ln\int\d z e^{-H_{\rm f,i}(z)/T}$ are free
energies referring to the initial temperature $T$, does not in general
provide any information about the principle, since for finite systems the
adiabatic work is not equal to $F_{\rm
f}-F_{\rm i}$; see \cite{minima,sato} for examples, and also below.

{\bf 1.} The initial
distribution ${\cal P}_{\rm i}(z)$ can be generated by sampling
microcanonical distribution ${\cal M}$ with initial energy probability 
$P_{\rm\, i}(E)$, 
\begin{gather}
{\cal M}(z,E,R_{\rm\, i})=\frac{1}{\omega_{\rm\,
i}(E)}\,\delta[E-H(z,R_{\rm\, i})],\\ 
P_{\rm\, i}(E)
=\frac{\omega_{\rm\, i}(E)}{Z_{\rm\, i}}e^{-\beta E},~~
\omega_{\rm\, i,f}(E)\equiv \int \d z\, \delta[E-H_{\rm\, i,f}(z)],\nonumber\\
\label{in}
{\cal P}_{\rm\, i}(z)=\int \d E \,P_{\rm\, i}(E)\,{\cal M}(z,E,R_{\rm\, i}).
\end{gather}
Hamilton's equations of motion imply $\frac{\d }{\d
t}H(z_t,R_t)=\dot{R}_t\,\partial_R H(z_t,R_t)$. 
On times $\tau_S\ll \tau\ll \tau_R$ we have for the energy change
\begin{gather}
\label{hek}
\Delta_\tau E
\equiv\frac{1}{\tau}[\,
H(z_{t+\tau},R_{t+\tau})-H(z_t,R_t)\,]=\\
\int_t^{t+\tau}\frac{\d s}{\tau}\,
\,\frac{\d H}{\d s}(z_s,R_s)
=\frac{\dot{R}_t}{\tau}\int_t^{t+\tau}\d s\,
\frac{\partial H}{\partial R}(z_s,R_t)+o(\frac{\tau}{\tau_R}).\nonumber
\end{gather}
The last integral refers to the frozen-parameter dynamics with $R_t=R$.
Now recall the Liouville theorem 
$\d z=\d z_t$ and energy conservation $H(z_{t+\tau})=H(z_t)=E_t$,
so that
\BEA
\label{karamba1}
\int \d z \,w(z){\cal M}(z,E_t)
=\frac{1}{\tau}\int_t^{t+\tau}\d s 
\int \d z \,w(z){\cal M}(z,E_t)\\
=\int \d z_t\,{\cal M}(z_t,E_t)\,
\frac{1}{\tau}\int_t^{t+\tau}\d s \,w(z[s;z_t]),
\label{karamba2}
\EEA
where $z[s;z_t]$ is the trajectory at time $s$, that starts at $z_t$ at time $t$.
If $w(z,R)=\partial_R H(z,R)$ is an {\it ergodic
observable} of the $R=$const dynamics, then
for $\tau\gg \tau_S$ the time-average in (\ref{karamba2}) does not
depend on the initial condition $z_t$ \cite{vk}, so the integration over 
$z_t$ in (\ref{karamba2}) is trivial, and we get from (\ref{karamba1})
that the time-average in (\ref{hek})
is equal to the microcanonical average at the energy $E_t$:
$\Delta_\tau E=\dot{R}_t \int \d z\,\partial_R H(z,R_t)\,{\cal M}(z,E_t,R_t)$.
This implies conservation of the phase-space volume $\Omega$:
\BEA
\label{der}
\Delta_\tau 
\Omega(E,R)=0,\,\,
\Omega(E,R)\equiv\int \d z\,\theta(E-H(z,R)).
\EEA
Given the initial energy $E_{\rm i}$ and the initial and final
values of $R_t$, $E_{\rm f}$ is found from $\Omega(E_{\rm
i}, R_{\rm i})= \Omega(E_{\rm f}, R_{\rm f})$. Since
$\partial_E\Omega(E,R)\equiv \omega(E,R)\geq 0$, we can define for fixed $R_{\rm f}$
and $R_{\rm i}$ two monotonous functions: $E_{\rm f}=\phi_{\rm f} (E_{\rm
i})$ and its inverse $E_{\rm i}=\phi_{\rm i} (E_{\rm f})$.  Note that
$\phi'_{\rm f}(E)=\omega_{\rm i}(E)/\omega_{\rm f}(\phi_{\rm f}(E))\geq 0$. 
The conservation of the phase-space volume for ergodic systems is known
for almost a century \cite{hertz}.  It led to the development of
microcanonical thermodynamics with entropy defined as $\ln \Omega$
\cite{berdi}. The precision of the above conservation was studied
in \cite{ott}. 

{\bf 2.} Let $P(E|E')$ be the conditional probability
of having energies $E$ and $E'$ at $t=t_{\rm f}$ and $t=t_{\rm i}$, 
respectively:
\begin{gather}
\label{get}
P(E|E')P_{\rm\, i}(E') = ~~~~~~~~~~~~~~~~~~~~~~~~~~~~~~~~~~
\\
\int \d z\d z'\delta[E-H_{\rm f}(z)   ]
\delta[E'-H_{\rm i}(z')   ]{\cal P}(z|z'){\cal P}_{\rm\, i}(z'),\nonumber
\end{gather}
where ${\cal P}(z|z')$ is the phase-space conditional probability
\BEA
{\cal P}(z|z')
={\cal U}\,\delta[p-p']\,\delta[q-q'], ~~
{\cal U}\equiv
\overleftarrow{e}^{\int_{t_{\rm i}}^{t_{\rm f}}\d s {\cal L}(s)},
\EEA
and where ${\cal L}(t)=\partial_q
H(t)\,\partial_p-\partial_pH(t)\,\partial_q$ is the Liouville operator, with
$\overleftarrow{e}$ being the chronological exponent.  We see that
$\int\d z' {\cal P}(z|z')=1$, since 
${\cal U}$ is a sum of $1$ and differential operators. This implies together 
with (\ref{in}, \ref{get}) 
\BEA
P(E|E')
&=&\int \d z\d z'\delta[E-H_{\rm f}(z)   ]
\frac{\delta[E'-H_{\rm i}(z_0)   ]}{\omega_{\rm\, i}(E')}{\cal P}(z|z'),\nonumber\\
\label{dd}
&&\int \d E' \omega_{\rm\, i}(E')   P(E|E') =\omega_{\rm f}(E),~
\EEA
in addition to the normalization $\int \d E P(E|E')=1$. 
For the adiabatic situation we get from conservation of $\Omega$:
\BEA
\label{kapri}
\widetilde{P}(E|E')=\delta[E-\phi_{\rm f}(E')].
\EEA

{\bf 3.} Now recall (\ref{wo}) and write $\Delta W$ as
\begin{gather}
\label{ko}
\Delta W=
\int \d z\,H_{\rm f}(z)[{\cal P}_{\rm\, f}(z) -\widetilde{{\cal P}}_{\rm\, f}(z) ]=\\
\int E \,\d E \int \d E' P_{\rm\, i}(E')[
P(E|E')- \widetilde{P}(E|E')]\equiv
\int \d E\,  g_E.\nonumber
\end{gather}
Integrating by parts, denoting $c_E(E')\equiv\int^E\d u P(u|E')$,
using (\ref{kapri}), and employing $\theta[E-\phi_{\rm\,f}(E')]=
\theta[\phi_{\rm\,i}(E)-E']$, 
\begin{gather}
g_E=\int^E\d u \int \d E'\, P_{\rm\, i}(E')\,[\,
\widetilde{P}(u|E')-P(u|E')\,]\nonumber\\
={\int}^{\phi_{\rm i}(E)}\d E' P_{\rm\, i}(E')
-\int \d E' P_{\rm\, i}(E')c_E(E')=\nonumber\\
\int^{\phi_{\rm i}(E)}\d E'P_{\rm i}(E')[1-c_E(E')]
-\int_{\phi_{\rm\, i}(E)}\d E'P_{\rm\, i}(E')c_E(E').\nonumber
\end{gather}
We now employ (\ref{in}) and then (\ref{dd}) to obtain
\begin{gather}
{\cal Z}_{\rm i}\,
g_E\geq 
e^{-\beta \phi_{\rm i}(E)}\int^{\phi_{\rm i}(E)}\d E'\omega_{\rm i}(E')[1-c_E(E')]
\nonumber\\
-e^{-\beta \phi_{\rm i}(E)}\int_{\phi_{\rm i}(E)}\d E'\omega_{\rm i}(E')c_E(E')\nonumber\\
=e^{-\beta \phi_{\rm i}(E)}\left[
\int^{\phi_{\rm i}(E)}\d E'\omega_{\rm i}(E')-
\int^{E}\d E'\omega_{\rm f}(E')
\right].
\label{kapo}
\end{gather}
Recalling that $\omega(E)=\Omega'(E)$ and that $\Omega(E_{\rm min})=0$ for
the lowest energy $E_{\rm min}$, we get (\ref{kapo})$=0$, i.e.,
$g_E\geq 0$.  This means, from (\ref{ko}), that the principle is
proven.  Note that the same proof applies for $P_{\rm\, i}(E)/\omega_{\rm
i}(E)$ being a monotonically decaying function of $E$. Neither the
proof, nor the principle itself, applies to the initial microcanonic
situation, where $P_{\rm\, i}(E)=\delta[E-E_{\rm i}]$ \cite{ANPhysicaA}. 

For {\it non-ergodic systems}, where under driving the system can move
from one ergodic component (of the $R=$const dynamics) to
another, the above proof of the principle is endangered, since in
general $\Omega$ in (\ref{der}) is not conserved. Indeed, the
argument expressed by (\ref{karamba1}, \ref{karamba2}) may not apply,
since now the time-average in (\ref{karamba2}) depends on the ergodic
component to which the initial condition $z_t$ belongs, and in general
it cannot be substituted by the microcanonical average over the full
phase-space.  However, $\partial_R H(z,R)$ may be ergodic, even though
the system is not \cite{vk}.  For an example consider the {\it
symmetric} double-well: $H=\frac{1}{2}{p^2}-R_t x^2+ g x^4$, with $g>0$.
For $E<0$, there are two ergodic components related by the inversion
$x\to -x$, but $\partial_R H(z,R)=-x^2$ is degenerate with respect to
them.  Though the $R_t=$const motion on the separatrix $E=0$ has an
infinite period (due to unstable fixed point $q=0$), when the initial
distribution is microcanonical, the fraction of particles 
trapped by the separatrix is neglegible \cite{best}, so that for
the ensemble $\tau_S$ remains finite.  Thus $\Omega(E,R)$ (with the
integration in (\ref{der}) over the whole phase-space) is conserved
\cite{best}.  Thus for this case the above proof of the principle applies. 

\begin{figure}
\includegraphics[width=6cm]{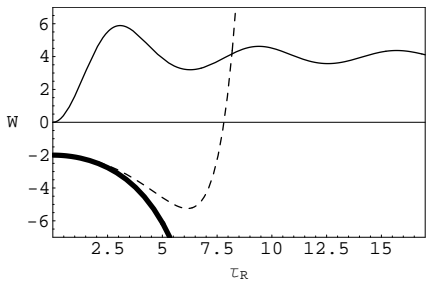} 
\vspace{0.05cm} 
\caption{Work $W$ versus the time-scale $\tau_R$ for an ensemble of harmonic 
oscillators with $R_t=-b+{t^2}/{\tau_R^2}$ at initial temperature $T=2$.
Full line: $b=0$, $t_{\rm f}=-t_{\rm f}=\tau_R$; the adiabatic work is $\widetilde{W}=4$,
as given by Eq.~(\ref{jj}). 
Dashed line: $b=0.5$, $t_{\rm i}=-\tau_R$ and $t_{\rm f}=\tau_R\sqrt{0.1}$.
Bold line: $b=0.5$, $t_{\rm i}=-\tau_R$ and $t_{\rm f}=0$. For the last two cases the adiabatic work
(the limit $\tau_R\to\infty$) is, respectively, plus and minus infinity.
The minimal work principle, $W(\tau_R)\ge W(\infty)$ for all $\tau_R$, 
is seen hold in the third case, but not in the first and second case. 
}
\label{fig_1} 
\end{figure} 

{\it Limits of the principle.} The above derivation of the principle
assumes that the frozen-parameter dynamics supports the microcanonical
distribution. This need not be always the case. Consider the basic model of the parametric
oscillator: $H=\frac{1}{2}p^2+\frac{1}{2}R_t q^2$. If $R_t$
is always positive, the phase-space volume is conserved, and the above
construction applies. But what happens if $R_t$ touches zero at one instant? 
This is another non-ergodic example, since for $R_t=0$ the frozen-parameter
phase-space consists of two ergodic components with, respectively, $p>0$
and $p<0$. Though $\partial_R H(z,R)=x^2/2$ is degenerate with respect to them,
the microcanonical distribution does not exist for $R_t=0$.

To separate the effect of initial conditions, we 
write the solution of the equations of motion 
$\ddot{q}+R_tq=0$ as
\BEA
\label{ka}
p(t)=\theta_{pp}\,p_{\rm i}+\theta_{pq}\,q_{\rm i}, 
~q(t)=\theta_{qq}\,q_{\rm i}+\theta_{qp}\,p_{\rm i},
\EEA
where $\theta_{kl}=\theta_{kl}(t)$.
With Gibbsian initial distribution 
${\cal P}_{\rm i}(p,q)\propto\exp[-\frac{p^2}{2T}-\frac{R_{\rm i}q^2}{2T}]$,
the work reads from (\ref{wo}):
\BEA
\label{kon}
\frac{W}{T}=-1+\half\theta^2_{pp}+\frac{\theta^2_{pq}}{2R_{\rm i}}
+\half R_{\rm f}\left(\theta^2_{qp}+\frac{\theta^2_{qq}}{R_{\rm i}}\right).
\EEA
Next we consider an exactly solvable situation $R_t=t^2/\tau_R^2$
(see below for generalizations).  The equation of motion is solved by
substitution $q(t)=\sqrt{t}\,x(t)$:
\BEA
\label{ne}
q(t)=
c_1 \sqrt{|t|}\,J_{-1/4}(\frac{t^2}{2\tau_R})+
c_2 \frac{t}{\sqrt{|t|}}\,J_{1/4}(\frac{t^2}{2\tau_R}),
\EEA
where $J_{\pm 1/4}$ are the Bessel functions, $c_1$ and $c_2$ are to be
found from initial conditions, and where $q(t)$ is written in a way that
applies to $t<0$: noting
$J_{\mu}(x\to 0^+)=\frac{x^\mu}{2^\mu\Gamma(\mu+1)}$ \cite{olver}, we see that 
$q(t)\approx \tilde c_1+\tilde c_2t$ near $t=0$. 
Let us define $t_{\rm i}=-\tau_R\sqrt{R_{\rm i}}$ and $t_{\rm f}=\tau_R\sqrt{R_{\rm f}}$.
In the slow limit $\tau_R\gg 1$ we need 
$J_{\mu}(x)=\sqrt{\frac{2}{\pi x}}\cos\left(
x-\frac{\pi\mu}{2}-\frac{\pi}{4}\right)+{\cal O}(x^{-3/2})$ \cite{olver}.
Eqs.~(\ref{ka}, \ref{ne}) then imply
\BEA
\label{qq1}
&&\theta_{qq}=R^{1/4}_{\rm i}\,R^{-1/4}_{\rm f}(-\sin u+\sqrt{2}\cos v),\\
&&\theta_{qp}=R^{-1/4}_{\rm i}\,R^{-1/4}_{\rm f}(\,\cos u+\sqrt{2}\sin v),\\
&&\theta_{pq}=R^{1/4}_{\rm i}\,R^{1/4}_{\rm f}(\,\cos u-\sqrt{2}\sin v),\\
&&\theta_{pp}=R^{-1/4}_{\rm i}\,R^{1/4}_{\rm f}(\,\sin u+\sqrt{2}\cos v),
\label{qq2}
\EEA
where $u=\half\tau_R(\sqrt{R_{\rm i}}-\sqrt{R_{\rm f}})$,
$v=\half \tau_R(\sqrt{R_{\rm i}}+\sqrt{R_{\rm f}})$.
Inserting (\ref{qq1}--\ref{qq2}) into (\ref{kon}) we get for the adiabatic work
\BEA
\label{jj}
\widetilde{W}=T(-1+3\sqrt{R_{\rm f}/R_{\rm i}}).
\EEA
Note that if $R_t$ is always positive, the conservation of the
phase-space volume $\Omega\propto E(t)R_t^{-1/2}$ gives
${\widetilde{W}}=\langle E_{\rm\, f}-E_{\rm\, i}\rangle=
(\sqrt{{R_{\rm f}}/{R_{\rm i}}}-1)\langle E_{\rm\, i}\rangle
=T(\sqrt{{R_{\rm f}}/{R_{\rm i}}}-1)$. 
Thus $\Omega$ is not conserved, if $R_t=0$
at one instant. Note that the adiabatic limit still exists, since
for a large $\tau_R$, $W$ converges to $\widetilde{W}$; see Fig.~\ref{fig_1}.
Eq.~(\ref{jj}) shows that the
minimal work principle does not hold.  Indeed, for a sudden variation
the Hamiltonian changes, while the state of the ensemble does not.
This brings
\BEA
\label{gog}
W_{\rm s}
=\int \d z{\cal P}_{\rm i}(z)[ H_{\rm f}(z)-H_{\rm i}(z)]
=T\left(\frac{R_{\rm f}}{2R_{\rm
i}}-\frac{1}{2}\right).
\EEA
$W_{\rm s}$ is sometimes smaller than (\ref{jj}), e.g., take $R_{\rm
i}=R_{\rm f}$; see also Fig.~\ref{fig_1}.  A qualitative picture behind
this is that for $R_t=0$ the particle runs to infinity, and to confine it
back (for $R_{\rm f}>0$), the work to be spent is larger for the slow
case, since for a quick process the particle does not have time to move
very far; see (\ref{gog}). 

An extended setup $R_t=-b+{t^2}/{\tau_R^2}$, $b\geq 0$, can be worked
out with help of hypergeometric functions.  Fig.~\ref{fig_1} illustrates
that the principle is satisfied if $R_t$ decays monotonically: $t_{\rm
i}<t_{\rm f}\leq 0$.  It is violated if the change of $R_t$ is
non-monotonic: $t_{\rm i}<0<t_{\rm f}$. The work does not saturate for
$\tau_R\to\infty$ if $R_t$ becomes strictly negative.  We checked
numerically that {\it i)} if $R_t$ is positive, but becomes very
small, $R=\delta$, the system has two widely different characteristic
times $\tau_S\sim 1$ and $\tau_S^*$ that diverges for $\delta\to 0$.
Thus there are two senses of being slow: the principle is limited for
$\tau_S\ll\tau_R\ll\tau_S^*$, while it is satisfied if
$\tau_S^*\ll\tau_R$, since then the phase-space volume is conserved, as
discussed above. {\it ii)} The limits of the principle exist for any
potential $U(q,R_t)$ that is globally confining ($U(q)\to \infty$ for
$q\to \pm \infty$) but looses this feature for one value of $R_t$.
These limits obviously exist for uncoupled particles.  We expect that
they extend to coupled particles put in a (de)confining potential. 

Note that whenever the principle gets limited via the above scenario, 
the slowest process is irreversible.  Recall that a
process is reversible, if after supplementing it with its mirror
reflection (the same process moved backwards with the same speed), the
work done for the total cyclic process is zero \cite{thermobooks}. As
seen from (\ref{jj}), the work  (equal to $2T$) does not vanish for the
cyclic adiabatic process with $R_t$ touching zero. Thus the process is
irreversible.  This fact contrasts the quantum limits of the minimal
work principle found in \cite{minima}.  Those limits are related to
energy level-crossing, where the adiabatic work is reversible \cite{minima}. 

Here is an experimentally realizable example that can
demonstrate the above limits. The simplest LC circuit consists of
capacitance $C$ and inductance $L$ (the resistance is either small or
compensated) \cite{chu}. The Hamiltonian is
$H=\frac{\Phi^2}{2L}+\frac{Q^2}{2C}$, where $Q$ (coordinate) is the
charge, and where $\Phi=L\frac{\d Q}{\d t}$ (momentum) is the magnetic flux.  
The parametric oscillator with $R_t\to
0$ corresponds to a time-dependent $C$ (or $L$) becoming very large at
some time. $R_t$ becoming negative at some time,
can be achieved via a {\it negative capacitance} $C_a<0$ given by a
special active circuit \cite{bu}. If such a capacitance is
sequentially added to a positive capacitance $C_n$, then the resulting
inverse capacitance $C^{-1}=C^{-1}_{a}+C^{-1}_{n}$ can be made zero and
then negative by tuning $C_n$. The same effect is obtained via a negative
inductance \cite{japan} added in parallel to a normal one. As the
negative inductance/capacitance emulators are widely appplied in
compensation of parasitic processes and for improving the radiation
pattern in antennas, \cite{chu,bu,japan} 
they can serve to test our predictions.

{\it In conclusion}, we studied the second law in its minimal work formulation 
for classical Hamiltonian systems. It was shown to hold under the assumption
that the observable of work (i.e., the derivative of the Hamiltonian w.r.t. the driven
parameter) is an ergodic function. The result applies to small systems.
There are, however, numerous examples of non-ergodicity both for finite
and macroscopic systems. For such systems we explored several
possiblities met in the one degree of freedom situation. The minimal
work principle applies if the observable of work is degenerate over
ergodic components and if the microcanonic equilibrium exists for all
values of the driven parameters. If the latter condition is not met, the
principle gets limits. The simplest example of the latter is provided
by a parametrically driven harmonic oscillator whose frequency passes through
zero.  As we saw, this situation can be realized experimentally in LC electrical circuits.
Multi-dimensional systems provide more complex examples of
non-ergodicity.  The understanding of the second law for such systems 
still deserves to be deepened, in view of the importance of non-ergodicity in
processes of measurement and information storage \cite{mi}. 

A.E. A. was supported by CRDF grant ARP2-2647-YE-05 and partially by FOM/NWO.


\begin{thebibliography}{99}

\bibitem{thermobooks} R. Balian, {\it From Microphysics to
Macrophysics}, volume I (Springer, 1992).  G. Lindblad, {\it
Non-Equilibrium Entropy and Irreversibility} (D.  Reidel, Dordrecht,
1983). 


\bibitem{venturi}
V. Balzani, A. Credi and M. Venturi, {\it Molecular
Devices and Machines} (Wiley-VCH, Weinheim, 2003).



\bibitem{epstein} P.S. Epstein,{\it Textbook of Thermodynamics},
(Wiley \& Sons, New York, 1937).

\bibitem{kur} 
G.N. Bochkov and Yu.E. Kuzovlev, Sov. Phys. 
JETP, {\bf 45}, 125 (1977). C. Jarzynski, Phys. Rev. Lett. {\bf 78}, 2690 (1997). 
For a review see J. Kurchan, cond-mat/0511073.


\bibitem{deri}
H. Narnhofer and W. Thirring, Phys. Rev. A, {\bf 26}, 3646 (1982). 
L.P. Kadanoff and P.C. Martin, Ann. Phys., {\bf 24},
419 (1963).
D. Forster, {\it Hydrodynamics, Broken Symmetry, and
Correlation Functions}, (Benjamin, New York, 1983).
R. Fukuda, Prog. Theor. Phys. {\bf 77}, 825 (1987).
K. Sekimoto and S. Sasa, J. Phys. Soc. Jpn. {\bf 66}, 3326 (1997).


\bibitem{deri1}H. Tasaki, cond-mat/0009206. 

C. Maes and H. Tasaki, 
cond-mat/0511419.



\bibitem{minima} A. E. Allahverdyan and Th.M. Nieuwenhuizen, Phys.
Rev. E {\bf 71}, 046107 (2005).

\bibitem{sato}
K. Sato {\it et al.}, Phys. Rev. E {\bf 66}, 016119 (2002).






\bibitem{hertz}P. Hertz, Ann. Phys. (Leipzig) {\bf 33}, 225 (1910); {\it
ibid.} {\bf 33}, 537.  T. Kasuga, Proc. Jpn. Acad. {\bf 37}, 366 (1961). 

\bibitem{berdi}
R. Becker, {\it Theory of Heat} (Springer, New York, 1967).

V. L. Berdichevsky, {\it Thermodynamics of Chaos and Order}, 
(Addison Wesley Longman, Essex, England, 1997).

S. Sasa and T.S. Komatsu, Prog. Theor. Phys. {\bf 103}, 1 (2000).
H.H. Rugh, Phys. Rev. E {\bf 64}, 055101 (2001).

\bibitem{ott}E. Ott, Phys. Rev. Lett. {\bf 42}, 1628 (1979).

C. Jarzynski, Phys. Rev. Lett. {\bf 71}, 839 (1993).

\bibitem{vk}N. G. van Kampen, Physica {\bf 53}, 98 (1971).

\bibitem{best}R.W.B. Best, Physica {\bf 40}, 182 (1968).



\bibitem{olver}F.W.J. Olver, {\it Asymptotics and Special Functions}
(Academic Press, New York, 1974).

\bibitem{ANPhysicaA} A.E. Allahverdyan and Th.M. Nieuwenhuizen, Physica A {\bf 305}, 542
(2002).

\bibitem{chu}L. Chua, C.A. Desoer, and E.S. Kuh, {\it Linear and Nonlinear
circuits} (McGraw-Hill, New-York, 1987).

\bibitem{bu}A.G. Butkovskii and Yu.I. Samoilenko, {\it Control of Quantum Mechanical Processes}
(Springer, Berlin, 1991).

\bibitem{japan}K.L. Su, IEEE Journal of Solid State Circuits, {\bf 2}, 22 (1967).
H. Funato {\it et al.}, IEEE Trans. Power Electronics, {\bf 12}, 589 (1997).

\bibitem{mi} A.E. Allahverdyan, R. Balian and Th.M. Nieuwenhuizen, Europhys.
Lett., {\bf 61} 452, (2003). J. Parrondo, Chaos, {\bf 11}, 725 (2001). S. Ishioka
and N. Fuchikami, {\it ibid.} 734.


\end{thebibliography}
\end{document}